\documentclass[aps,prl,twocolumn,groupedaddress,floatfix,10pt]{revtex4}
\usepackage[pdftex]{graphicx,color}
\usepackage{dcolumn}
\usepackage{bm}

\usepackage{verbatim}
\usepackage{amsfonts}
\usepackage{amssymb}
\usepackage{amsmath}
\usepackage{amsthm}

\usepackage{scalerel,stackengine}

\usepackage{simpler-wick} 
\usepackage[caption=false]{subfig}
\usepackage{rotate}
\usepackage{dsfont}
\usepackage{hyperref}

\let\oldmarginpar\marginpar
\renewcommand\marginpar[1]{\-\oldmarginpar[\raggedleft\tiny #1]%
{\raggedright\tiny #1}}

\def\KeyWord#1{$\backslash$\IfColor{$\!\!$\textRed{#1}\textBlack}{#1}$\!\!$}

\newcommand{\ket}[1]{|#1\rangle}

\DeclareMathOperator{\Tr}{Tr}

\newcommand\equalhat{\mathrel{\stackon[1.5pt]{=}{\stretchto{%
    \scalerel*[\widthof{=}]{\wedge}{\rule{1ex}{3ex}}}{0.5ex}}}}

\begin{document}

\title{Universal entanglement of typical states in constrained systems}

\author{S. C. Morampudi}

\author{A. Chandran}

\author{C. R. Laumann}

\affiliation{Department of Physics, Boston University, Boston, MA 02215, USA}

\begin{abstract}
Local constraints play an important role in the effective description of many quantum systems.
Their impact on dynamics and entanglement thermalization are just beginning to be unravelled. 
We develop a large $N$ diagrammatic formalism to exactly evaluate the bipartite entanglement of random pure states in large constrained Hilbert spaces.
The resulting entanglement spectra may be classified into `phases' depending on their singularities.  
Our closed solution for the spectra in the simplest class of constraints reveals a non-trivial phase diagram with a Marchenko-Pastur (MP) phase which terminates in a critical point with new singularities.  
The much studied Rydberg-blockaded/Fibonacci chain lies in the MP phase with a modified Page correction to the entanglement entropy, $\Delta S_1 = 0.513595\cdots$. 
Our results predict the entanglement of infinite temperature eigenstates in thermalizing constrained systems and provide a baseline for numerical studies. 
\end{abstract}

\maketitle

Random matrix theory is widely used to describe statistical properties of quantum chaotic systems \cite{Wigner:1955aa,Dyson:1962p11206,Mehta:2004aa,Guhr:1998aa,Kravtsov:2009aa}.
%
Consider a bipartition of a random pure state $\ket{\psi}$ into two equal subsystems.
The entanglement across the cut is captured by the density of states (DOS) of the reduced density matrix $\hat{\rho}$ \cite{Haldane2008}.
Random matrix theory predicts that the DOS is given by the Marchenko-Pastur (MP) distribution \cite{Marcenko:1967aa},
\begin{align}
\label{eq:mplaw}
 p_{MP}(\epsilon) = \frac{1}{2\pi} \sqrt{\frac{4-\epsilon}{\epsilon}} \qquad \textrm{for } \epsilon \in (0,4).
\end{align}
The connection to quantum chaos is provided by the hypothesis that eigenstates and late-time states at infinite temperature are indistinguishable from random pure states within subsystems\cite{Deutsch1991, Srednicki1994,Rigol:2008aa,Goldstein:2006aa,Popescu:2006aa,DAlessio:2016aa,Yang:2015ab,Geraedts:2017aa}.
This hypothesis, along with Eq.~\eqref{eq:mplaw}, predicts the widely-observed Page correction \cite{Page1994} to the entanglement entropy of eigenstates \cite{D_Alessio_2014,Zhang:2015aa,Luitz:2015fj,Ponte:2015zp,Zhang:2016aa}.

Many physical systems, such as frustrated magnets\cite{Balents2010, Moessner2011}, pinned non-Abelian anyons\cite{Feiguin2007, Gils2009} and ultracold Rydberg ensembles\cite{Saffman:2010aa,Bernien2017}, inhabit locally constrained effective Hilbert spaces.
Recent measurements of entanglement\cite{Kaufman:2016aa} and quench dynamics in Rydberg-blockaded chains\cite{Bernien2017} raise questions about the structure of entanglement in constrained spaces \cite{Chandran2016, Lan2017, Chen2018, Brenes2018,Turner:2018aa}.
The lack of tensor product structure implies that even the reduced density matrix of the infinite temperature mixed state, $\hat{\rho}^{T=\infty}$, is not a simple identity.
We analytically compute the entanglement DOS $p(\epsilon)$ of random pure states for a wide array of constrained Hilbert spaces and show that, in general, $p(\epsilon)$ deviates from the MP law.
Our exact diagrammatic technique represents $p(\epsilon)$ in terms of diagrams which become planar in the thermodynamic limit.
The constraints dress the diagrams with nontrivial `spin' structure which fortunately still admit a resummation.

\begin{figure}[b]
    \centering
    \includegraphics[width=\columnwidth]{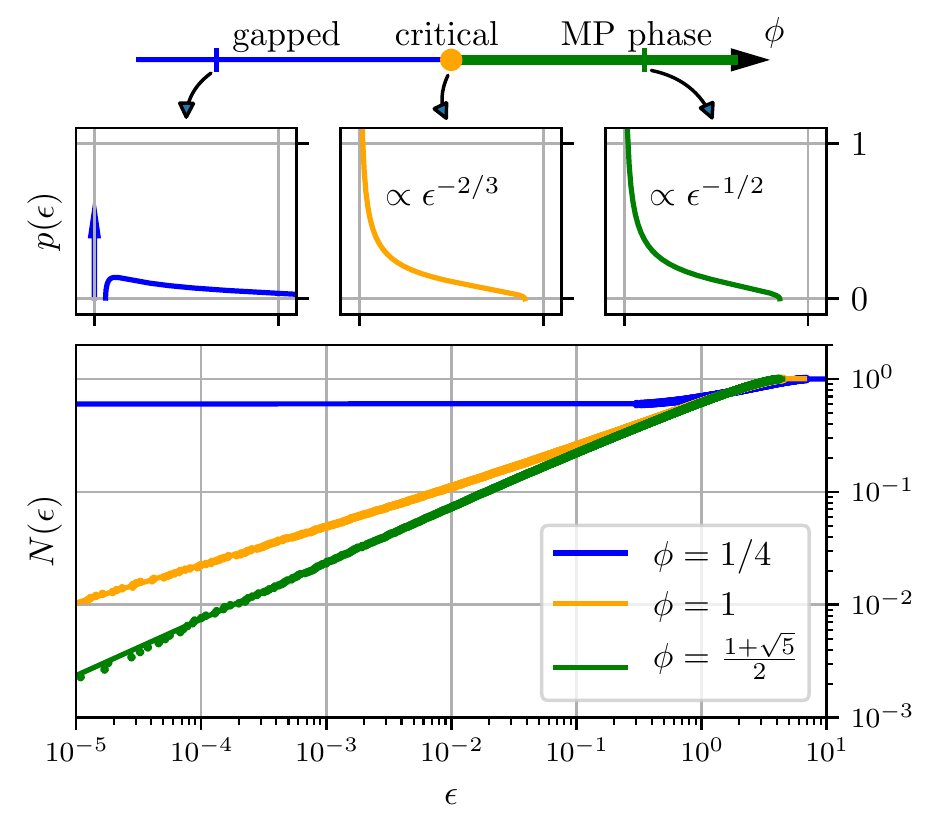}
    \caption{Evolution of the entanglement DOS $p(\epsilon)$ as a function of the relative dimension $\phi$ of the sectors $l=0,1$. 
    The critical divergence $p \sim \epsilon^{-1/2}$ persists through the MP phase ($\phi > \phi_c = 1$, green). 
    This gives way to a gapped spectrum with a finite weight delta function at $\epsilon=0$  ($\phi < 1$, blue) across a multicritical point ($\phi=1$, orange) with distinct critical exponents, $p \sim \epsilon^{-2/3}$. 
    (bottom) Integrated density of states $N(\epsilon)$ at three representative values, including the Rydberg/Fibonacci chain at $\phi = (1+\sqrt{5})/2$. Solid lines are analytic forms with numerical data from the diagonalization of an $N=1000$ random pure state overlaid (points). 
    }
    \label{fig:dos_constrained}
\end{figure}

The entanglement distribution $p(\epsilon)$ is `universal' in two senses.
First, for a given constrained space, we expect $p(\epsilon)$ to describe the entanglement of eigenstates and late-time states at infinite temperature of \emph{any} quantum chaotic Hamiltonian.
Second, $p(\epsilon)$ can be grouped into `entanglement phases' as a function of the constraint parameters according to their singularities at small $\epsilon$.
Fig.\ref{fig:dos_constrained} shows an example phase diagram.
Remarkably, the $\epsilon^{-1/2}$ singularity of the MP distribution at the unconstrained point $\phi = \infty$ is stable to the addition of constraints, extending down to $\phi=1$.
The singularity is modified at this new critical point to $\epsilon^{-2/3}$.

Below we discuss how to parametrize constrained Hilbert spaces, before turning to the diagrammatic method.
We then solve for the entanglement DOS in two cases.
First, we consider a locally constrained family of models which includes the Rydberg/Fibonacci chain and derive Fig.~\ref{fig:dos_constrained}. For the special case of the Rydberg chain, we compute the Page corrections to the von Neumann and Renyi entropies. Second, we consider systems with global conservation laws which produce diagonal constraints. We close with general comments on entanglement phases and the Page correction.

\begin{figure}[tb]
	\centering
	\includegraphics[width=\columnwidth]{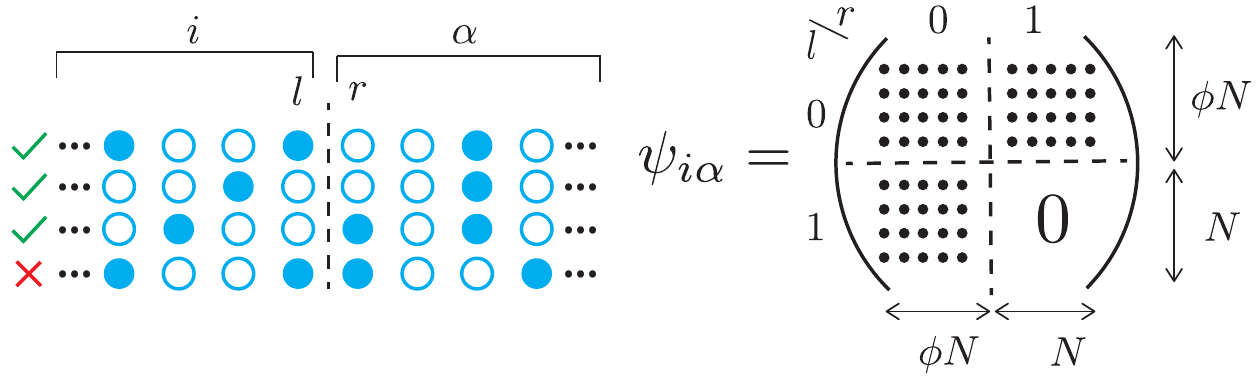}
	\caption{
	(left) Configurations of a blockaded chain. 
	The bottom configuration is disallowed by the constraint. 
	The multi-indices $i$ and $\alpha$ label configurations in the left and right subsystems, while $l,r$ label the states of the boundary sites. 
	(right) 
	Block structure of a pure state. The constraints impose that the $(1,1)$ block is zero.
	}
	\label{fig:chain}
\end{figure}

\paragraph{Parametrizing constrained spaces---}
Our primary example of a constrained system is a `blockaded' chain. 
Each site of the chain may be in state $0$ or $1$, subject to the constraint that no two consecutive sites are $1$ (see Fig.~\ref{fig:chain}). 
The blockaded Hilbert space arises naturally in certain Rydberg atom experiments \cite{Bernien2017}, and in chains of pinned Fibonacci anyons in the global $\tau$ fusion channel \cite{Feiguin2007, Gils2009}.
We will use the language of the Rydberg chain for concreteness.

Consider a length $2L$ chain which we partition in two equal halves.
The states of the left half can be grouped into two boundary condition sectors labeled by $l=0,1$ corresponding to the state of the $L$'th site.
The dimension of the $l$-sector is $D_l = F_{L+1-l}$, where $F_L$ is the $L$'th Fibonacci number.
At large $L$, we have that $D_0 = \phi D_1 \equiv \phi N$ where the relative dimension, $\phi = (1+\sqrt{5})/2$, is the golden mean, and $N \sim \phi^L$.
Similarly, the right half chain has two sectors labeled by $r=0,1$, corresponding to the state of the $(L+1)$'th site.
The constraint across the cut imposes that $(l,r) \ne (1,1)$. 

At large $N$, the entanglement across the cut of a random pure state depends only on the constraint across the cut and the relative dimensions of the different boundary condition sectors. 
We parameterize the constraint by a matrix $C$ with $C_{lr}=1$ if $lr$ is allowed and $0$ otherwise. 
The relative dimensions are represented by vectors $d_{l} \equiv D_{l}/N$ and $d'_r \equiv D_{r}/N$. For the blockaded chain,
\begin{align}
    \label{eq:CD_fibo}
	C&=\begin{pmatrix}
		1&1\\
		1&0
	\end{pmatrix}& d&=d'=\begin{pmatrix} \phi \\ 1 \end{pmatrix}.
\end{align}
In terms of $C$, $d$ and $d'$, our formalism generalizes to a wide array of constrained systems. 
Note that $C=(1),\, d=d'=(1)$ in an unconstrained system.
Much of the discussion below is general; as a running example, we specialize formulae to the blockaded chain using $\equalhat$.

\paragraph{Constrained diagrams---}
Let $|\psi\rangle$ be a Gaussian random pure state normalized such that
\begin{align}
\label{eq:psi_psidagger_avg}
	\overline{\psi_{i\alpha}\psi^\dagger_{\beta j}}& = \frac{1}{N}\delta_{i\alpha,j\beta} = \frac{1}{N}\sum_{l,r} C_{lr} \delta^l_{ij}\delta^r_{\alpha\beta} \\
	&\equalhat \frac{1}{N} \left( \delta^0_{ij} \delta^0_{\alpha\beta} + \delta^1_{ij} \delta^0_{\alpha\beta} + \delta^0_{ij} \delta^1_{\alpha\beta}\right) \nonumber
\end{align}
Here, $\psi^\dagger_{\beta j} = \psi^*_{j \beta}$ is the conjugate transpose of the amplitudes of $|\psi\rangle$ viewed as a matrix in the left ($i,j$) and right ($\alpha,\beta$) indices, and $\overline{~\cdot~}$ denotes averaging over the Gaussian ensemble.
The operators $\delta^l_{ij}$ and $\delta^r_{\alpha\beta}$ project onto the $l$ and $r$ boundary condition sectors of the left and right subsystems, respectively. 
That is, $\delta^l_{ij}=1$ if $i = j$ and $i$ has boundary spin value $l$. 
With this notation, the reduced density matrix for the left subsystem is $\hat{\rho} = \psi \psi^\dagger$.

The normalization of Eq.~\eqref{eq:psi_psidagger_avg} is convenient for the diagrammatics below. 
However, one must appropriately include normalization factors of
\begin{align}
\label{eq:norm}
	\mathcal{N} = \overline{\Tr \hat{\rho}} =  N \sum_{lr} C_{lr} d_l d'_r \equalhat N(\phi^2 + 2\phi)
\end{align}
in final formulae.

The calculation of the ensemble averaged trace moments of the reduced density operator, $ \overline{\Tr\hat{\rho}^n}$, may be organized diagrammatically using Wick's theorem as follows.
(1) Introduce a single solid/dashed line to represent each left ($i,j$)/right ($\alpha,\beta$) index contraction.
\begin{align}
	\delta^l_{ij} &= \vcenter{\hbox{\includegraphics{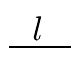}}} 
	& \delta^r_{\alpha\beta} &= \vcenter{\hbox{\includegraphics{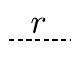}}} 
\end{align} 
Each line carries a `spin' label $l$/$r$ which indicates the boundary sector and which must be summed over to evaluate the diagram.
(2) Introduce vertices for each insertion of $\psi$ or $\psi^\dagger$:
\begin{align}
	\psi_{i\alpha} &= \vcenter{\hbox{\includegraphics{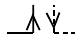}}} 
	& \psi^\dagger_{\beta j} &= \vcenter{\hbox{\includegraphics{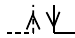}}} 
\end{align}
In $\Tr \hat{\rho}^n$, the $\psi$ and $\psi^\dagger$ factors alternate so these two rules produce a circle of alternating solid and dashed lines. 
Finally, (3) each Wick pairing is represented by a double line,
\begin{align}
	\wick{ \c1 \psi_{i\alpha} \c1 \psi^{\dagger}_{\beta j}} &= \vcenter{\hbox{\includegraphics{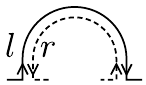}}}
\end{align}
connecting the tops of the vertices consistent with the arrows and solid/dashed lines. 
The spin labels carried by the double lines must satisfy the constraint encoded by $C$.

For example, the second trace moment may be evaluated diagrammatically,
\begin{align}
\label{eq:renyi2_diags}
	\Tr \overline{\hat{\rho}^2} &= \vcenter{\hbox{\includegraphics{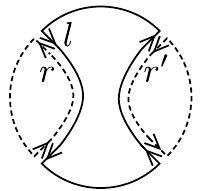}}} + \vcenter{\hbox{\includegraphics{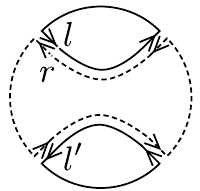}}} \nonumber\\
	&\equalhat 2N(\phi^3 + 3 \phi^2 + \phi)
\end{align}

Each diagram arising in the evaluation of $\Tr \overline{\hat{\rho}^n}$ may be viewed as a surface with a circular boundary~\cite{Hooft:1974aa,Brezin:1994aa,Brouwer:1996aa,Jurkiewicz:2008aa,Zee:2010aa,Morampudi:2018aa}. The double lines are the internal edges separating faces (loops) while the single lines make up the outer boundary.
Since each closed loop contributes a factor of $N d_l$ or $N d'_r$ and each double line contributes $1/N$, the $N$ dependence of the diagram is given by the Euler characteristic $\chi$ of the surface. 
Accordingly, the large $N$ limit picks out the planar diagrams with $\chi = 1$.
Moreover, the relative fluctuations of any trace moment vanish at large $N$: the connected correlators, $\overline{(\Tr \cdot )(\Tr \cdot)}_c$, have two circular boundaries so that $\chi \le 0$.
The entire sequence of trace moments and corresponding DOS is thus self-averaging at large $N$. 

In general, there are $C_n$ planar diagrams in $\Tr \overline{\hat{\rho}^n}$, each of which contributes a non-trivial polynomial in the relative dimensions.
Here, $C_n$ is the $n$'th Catalan number.
For the unconstrained case, the polynomial is trivial and $\Tr \overline{\hat{\rho}^n} = N C_n$ correctly reproduces the moments of the MP law.

\paragraph{Entanglement spectrum---}
To calculate the density of states (DOS) of $\hat\rho$, we turn to the resolvent 
\begin{align}
	\hat{G}(z) &= \overline{\frac{1}{z-\hat{\rho}}} 
\end{align}
which, by the Haar symmetry of $|\psi\rangle$, is diagonal and constant in each sector $l$ of the left Hilbert space,
\begin{align}
    \hat{G}(z) &= \sum_l \delta^l G_l(z) \equalhat \delta^0 G_0(z) + \delta^1 G_1(z)
\end{align}
The diagonal matrix elements  $G_l(z)$ determine the mean entanglement DOS associated with the $l$ sector,
\begin{align}
    p_l(x) = \frac{-1}{\pi} \mathrm{Im~} G_l(x+i0^+)
\end{align}
The distribution $p_l$ is normalized so that $\int dx\, p_l(x) = 1$; the total DOS takes into account the relative dimension between the sectors:
\begin{align}
    p(x) &= \frac{\sum d_l p_l(x)}{\sum d_l} \equalhat \frac{\phi p_0(x) + p_1(x)}{\phi + 1}
\end{align}
In order to correct for the normalization of Eq.~\eqref{eq:norm}, we will eventually change variables from $x$ to
\begin{align}
    \epsilon = \frac{N \sum_l d_l}{\mathcal{N}} x \equalhat \frac{\phi+1}{\phi^2 + 2\phi} x
\end{align}
This scales the eigenvalues of $\hat{\rho}$ such that $\Tr \hat{\rho}$ is the dimension of the left Hilbert space.

To compute the resolvent, $\hat{G}(z)$, we expand,
\begin{align}
    \hat{G}(z) &= \frac{1}{z}\sum_{k=0}^\infty \frac{\overline{\hat{\rho}^k}}{z^k} 
\end{align}
and sum over all planar diagrams generated by the Wick contractions of $\overline{\hat{\rho}^k}=\overline{(\psi \psi^\dagger)^k}$. 
We rearrange the sum over diagrams to obtain a Dyson series with self energy relations of the following form,
\begin{align}
\vcenter{\hbox{\includegraphics{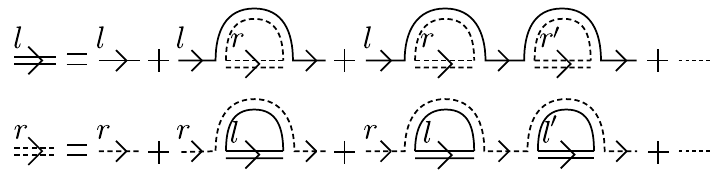}}}    
\end{align}
In these diagrams, the single solid line (bare propagator for left indices) carries an extra $1/z$ compared to the rules described in the previous section.

Resumming the Dyson series we obtain the governing equations,
\begin{align}
\label{eq:selfenergyrelations}
	G_l &= \frac{1}{z-\Sigma_l} &
	\Sigma_l &= \sum_r C_{lr} d'_r H_r \\
	H_r &= \frac{1}{1-\Sigma'_r} &
	\Sigma'_r &= \sum_l C_{lr} d_l G_l
\end{align}
where $H_r$ denotes the full propagator for the dashed lines and $\Sigma$ and $\Sigma'$ are self energies. 
The correct branch of solutions for $G_l(z)$ must be analytic in the upper half plane and decay as $1/z$ in order to correspond to a properly normalized $p_l(x)$.

As a simple check, consider the bipartition in an unconstrained space.
Eq.~\eqref{eq:selfenergyrelations} reduces to a quadratic equation for $G(z)$ whose solution, $ G_{MP}(z) = (1 - \sqrt{1-4/z})/2$,
indeed reproduces the MP law, Eq.~\eqref{eq:mplaw}.

\paragraph{Blockaded chain---}
For the constraint structure of Eq.~\eqref{eq:CD_fibo}, some algebra reveals that $G_0$ satisfies a cubic equation,
\begin{align}
\label{eq:g0_cubic_poly}
	G_0^3 -\frac{2}{\phi} G_0^2 + \frac{z+\phi-1}{z\phi^2} G_0 -\frac{1}{z\phi^2} = 0
\end{align}
and that $G_1$ can be expressed in terms of $G_0$,
\begin{align}
\label{eq:g1_from_g0}
	\frac{1}{G_1} &= \frac{1}{G_0} + \frac{1}{1-\phi G_0}.
\end{align}
The complete solution of these equations exploiting the Vieta formula can be found in the Appendix. 

The salient features of the resulting entanglement spectra are illustrated in Fig.~\ref{fig:dos_constrained}. 
As a function of the relative dimension $\phi$, there are three `entanglement phases', distinguished by the nature of the singularity in the DOS $p(\epsilon)$ as $\epsilon \to 0$. 
We dub the regime $\phi > 1$ the \emph{MP phase} as $p(\epsilon) \sim \epsilon^{-1/2}$, just like the MP law Eq.~\eqref{eq:mplaw}. 
Indeed, for $\phi \to \infty$, the $(0,0)$ sector of the composite Hilbert space dominates the amplitude matrix $\psi_{i\alpha}$ and $p(\epsilon)$ approaches the MP law exactly.
The support of the DOS, $(0, z_+)$, however changes smoothly with $\phi$. 
Note that the Rydberg/Fibonacci chain where $\phi = (1+\sqrt{5})/2$ lies in the MP phase.

For $\phi < 1$, the continuous part of the DOS gaps away from a delta function located at $\epsilon = 0$. 
The delta function has mass $\frac{1-\phi}{1+\phi}$, which follows from the number of linearly independent columns in a full rank matrix $\psi_{i\alpha}$ with the block structure shown in Fig.~\ref{fig:chain}.
Qualitatively similar behavior arises in unconstrained systems when the dimensions of the two subsystems are unequal.

The multicritical point at $\phi= 1$ has no counterpart in unconstrained systems. 
The DOS exhibits a non-trivial power law $p(\epsilon) \sim \epsilon^{-2/3}$. 

\paragraph{Entropy corrections---}
There is very little information regarding a global random pure state from measurements confined to a subsystem. 
This can be quantified using the Renyi and von Neumann entanglement entropies,
\begin{align}
    S_n(\hat{\rho}) \equiv \frac{1}{1-n}\ln \frac{\Tr \hat{\rho}^n}{(\Tr \hat{\rho})^n}.
\end{align}
A typical random pure state in an unconstrained system has entropy nearly that of the infinite temperature mixed state on the same system. 
Indeed, Page famously showed that the deviation
\begin{align}
\label{Eq:PageDef}
    \Delta S_n =  S_n\left(\hat{\rho}^{T=\infty}\right) - \overline{S_n(\hat{\rho})},
\end{align}
is less than or equal to $1/2$ for $n=1$ even though $S_1$ itself is extensive. 

We expect modifications to both terms in Eq.~\eqref{Eq:PageDef} for constrained systems because 
(i) \mbox{$\hat{\rho}^{T=\infty} \propto  \sum_{lr} \delta^l C_{lr} d'_r$} is not the identity operator on the left space, and 
(ii) the entanglement DOS $p(\epsilon)$ for the random state is different from the MP law.
As the fluctuations of trace moments vanish at large $N$ (i.e. there is concentration of measure), the average entropy depends explicitly on $p(\epsilon)$:
\begin{align}
    \overline{S_n(\hat{\rho})} &= \ln \left[ N\sum d_l\right] + \frac{1}{1-n}  \ln \left[\int d\epsilon p(\epsilon)\, \epsilon^n \right] 
\end{align}

\begin{figure}[tb]
	\centering
	\includegraphics[width=\columnwidth]{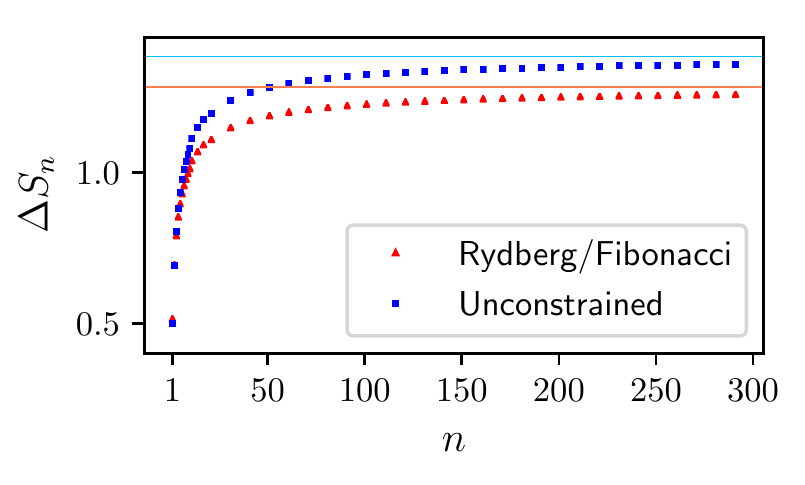}
	\caption{The Page corrections to the von Neumann ($n=1$) and Renyi entropies ($n \geq 2$) for the Rydberg/Fibonacci chain (red) and the unconstrained chain (blue). At large $n$, the Rydberg corrections approach the value $\ln z_+ - \ln[(\phi +1)^2/{\phi^2+2\phi}]$ (red line), while the unconstrained corrections approach $\ln 4$ (blue line).}
	\label{fig:entropies}
\end{figure}

Fig.~\ref{fig:entropies} shows the Page corrections for the Rydberg-blockaded ($\phi = (1+\sqrt{5})/2$) and unconstrained chain as a function of Renyi index $n$ (see Appendix for more details). 

Intriguingly, the Page correction at $n=1$ for the Rydberg chain:
 \begin{align}
    \Delta S_1 \approx 0.513595\cdots
\end{align}
is larger than the value $1/2$ of the unconstrained chain, violating the Page inequality. 
The half subsystem has more information regarding the pure global state in the constrained chain than it does in the unconstrained case.

\paragraph{Diagonal constraints---} 
Many systems have \emph{diagonal} constraints, which impose a one-to-one relationship between allowed $l$ and $r$ sectors.
For example, global $S^z$ conservation constrains the total $S^z$ on the left $l$ and right $r$ to add up to the total magnetization $M$.
Systems of pinned non-Abelian anyons provide a more exotic example.
Net fusion into the vacuum channel constrains left and right fusion sectors to be conjugate \cite{Bonderson:2007uh}. 

As each $(l,r)$ sector may be viewed as a product space, the total DOS is a linear combination of scaled and re-weighted MP laws:
\begin{align}
\label{eq:mp_linearcombo}
    p(x) &\propto \sum_l \frac{d_l}{d'_r} p_{MP}^{(\lambda_l)}\left (\frac{x}{ d'_r}\right).
\end{align}
Here, $p_{MP}^{(\lambda)}(x)$ is the unbalanced MP distribution, which arises when $\lambda_l = d_l / d'_r$ deviates from 1. 

In a spin-$1/2$ chain with $S^z$ conservation, $\lambda = \binom{L}{M-l} / \binom{L}{l}$ is generically not equal to 1.
Direct computation from \eqref{eq:mp_linearcombo} reproduces the Renyi entropies calculated in \cite{Lu:2017aa,Garrison:2015aa,Dymarsky:2018aa,Vidmar:2017aa} for globally constrained systems \footnote{We note that \cite{Lu:2017aa,Garrison:2015aa,Dymarsky:2018aa} take the Gibbs ensemble to be defined by a subsystem Hamiltonian rather than the partial trace of a global Gibbs state. The Page corrections following from our definition remain $O(1)$ for all Renyi indices $n$.}.

The balanced case, where all allowed sectors have $d_l = d'_r$, arises for the $S^z$-conserving chain at $M=0$ or for symmetric bipartitions of pinned anyons.
Although the total DOS $p(x) \ne p_{MP}(x)$, it lies in the MP phase since the $\epsilon^{-1/2}$ singularity persists.
Interestingly, the Page corrections are actually \emph{unmodified} from those of the simple MP law, $\Delta S_n = \frac{1}{n-1}\ln C_n$.

\paragraph{Discussion---}

We have developed a diagrammatic technique to compute the entanglement DOS $p(\epsilon)$ of random pure states in a wide array of constrained systems. 
This leads to entanglement `phases' which are classified by the singularity structure of $p(\epsilon)$ at small $\epsilon$.
We have focused on bipartitions into equally sized left and right subsystems. 
For unconstrained systems, it is well known that $p^{(\lambda)}_{MP}(\epsilon)$ gaps for unbalanced cuts. 
This extends to constrained systems.
Taking $d = \lambda d'$, the MP-phase in Fig.~\ref{fig:dos_constrained} may be viewed as a critical boundary at $\lambda = 1$ in the $\phi-\lambda$ plane. 
It has two natural critical exponents, $p(\epsilon) \sim \epsilon^{-1/2}$ and $\Delta \sim |\lambda-1|^2$, governing the DOS and gap scaling, respectively.
The multicritical point at $\phi=\lambda=1$ terminates the MP-phase with  exponents, $p(\epsilon)\sim \epsilon^{-2/3}$ and $\Delta \sim (1-\phi)^{3}$ (on the $\lambda =1$ boundary). 
The multicritical point is readily realized by an otherwise unconstrained spin-$1/2$ chain with an infinite penalty for the $(1,1)$ state of the two spins across the central cut.

We do not expect to be able to solve for the DOS for constrained systems with more than two boundary sectors in closed form: the self-consistency equations will be higher than quartic order. 
Nevertheless, the equations can be analyzed asymptotically at small $z$ in order to extract the entanglement phases and numerically solved to extract $p(\epsilon)$ to any desired precision.

The Page correction $\Delta S_1$ quantifies the information that half of a system has regarding the purity of the global state. 
Intriguingly, $\Delta S_1(\phi) > 1/2$ for all $\phi$; heuristically, the blockade constraint provides more information to the subsystem than is available in the unconstrained case.
We conjecture that Page's result is a lower bound to $\Delta S_1$ across all  constrained systems.

\begin{acknowledgments}
\paragraph{Acknowledgements---}
The authors would like to thank C. Chamon, A. Hamma, P. Krapivsky, A. Polkovnikov and Z. Yang for stimulating discussions. 
C.R.L. and A.C. acknowledge support from the Sloan Foundation through Sloan Research Fellowships, and from the NSF through grants PHY-1752727 and DMR-1752759, respectively.
Any opinion, findings, and conclusions or recommendations expressed in this material are those of the authors and do not necessarily reflect the views of the NSF.
\end{acknowledgments}

\bibliography{refs}

\appendix

\section{Resolvent of blockaded chain}
\label{app:solblockaded}

In this appendix, we include explicit solutions for the resolvent and entanglement DOS of the blockaded chain. 
We note that this section is expressed entirely in terms of $z = x+iy$ which needs to be rescaled by $(\phi+1) / (\phi^2+2\phi)$ in order to obtain correctly normalized eigenvalues of the reduced density matrix $\epsilon$.

The three roots of Eq.~\eqref{eq:g0_cubic_poly} can be expressed explicitly using Vieta's formula,
\begin{align}
\label{eq:g0_explicit}
	G_0 &= \frac{2}{3\phi} - \frac{1}{3} 
	\left[ e^{2\pi i k / 3} A^{1/3} + \frac{\Delta_0}{e^{2\pi i k /3} A^{1/3}}\right] \\
	A &= \frac{\Delta_1 - \sqrt{-27 \Delta}}{2} 
\end{align}
where $k=0,1,2$ and
\begin{align}
\label{eq:discriminants}
	\Delta &=\frac{4(z-z_-)(z-z+)}{z^3\phi^6} \\
	\Delta_0 &= \frac{1}{\phi^2} - \frac{3(\phi-1)}{z \phi^2} \\
	\Delta_1 &= \frac{2}{\phi^3} - \frac{9(2+\phi)}{z \phi^3} \\
	z_\pm &= \frac{1}{8} \left(- \phi^2 +20\phi +8  \pm \sqrt{\phi (\phi +8)^{3}} \right)
\end{align}
Here, $\Delta$ is the discriminant of Eq.~\eqref{eq:g0_cubic_poly} and $z_\pm$ are the zeros of $\Delta$ in the $z$-plane.  

We make a few general comments before turning to the three different cases of $\phi > 1$, $\phi =1$ and $\phi < 1$.

In general, the correct branch of $G_0$ must decay as $1/z$ at large $z$ and be analytic in the upper half plane.
At large $|z|$, these requirements pick the $k=0$ solution on using the principal branches of the square and cube roots in Eq.~\eqref{eq:g0_explicit}. 
Care must be taken in order to analytically continue this branch so that it remains smooth in the entire upper half plane.

The discriminant $\Delta$ helps to identify the support of the continuous part of the DOS $p_l(x) = -\frac{1}{\pi} \Im G_l(x+i0)$. 
On the real line, the imaginary part of $G_0$ can be non-zero \emph{only if} $\Delta < 0$. 
This immediately implies $p_l(x) = 0$ for $x > z_+$. 
The discriminant $\Delta$ changes sign at $z_\pm$ and $0$; as we will see below, the support of the spectrum is determined by the ordering of these sign changes.

Since $\hat{\rho}$ is a positive operator, we further know that $p_l(x) = 0$ for $x < 0$. This provides a check on our analytic continuation.

\paragraph{MP-phase ($\phi>1$)---} 
For $\phi>1$, $z_- < 0 < z_+$, and the discriminant $\Delta < 0$ for (i) $z < z_-$ and (ii) $0 < z < z_+$.

In region (i), $\Delta_1 > \sqrt{-27 \Delta}$ so $A$ is positive and the $k=0$ solution encounters no branch points following from $z = -\infty$. 
This implies $G_0$ follows the real branch, consistent with the lack of DOS on the negative axis.

In region (ii), the correct branch of $G_0$ is the complex solution with negative imaginary part. 
This can be tediously understood by tracing the singularities and zeros in the Vieta expressions above. 

At the end of the day, the mean spectrum in the zero sector has support on the interval $[0,z_+]$, and is given by
\begin{align}
\label{eq:p0_mp_explicit}
	p_0(x) 
	&=  \frac{1}{3\pi} \sin\left(\frac{\pi}{3}\right) \left( |A(x)|^{1/3} - \frac{\Delta_0(x)}{|A(x)|^{1/3}}\right)
\end{align}
where $A \in \mathbb{R}$ is evaluated on the real line at $z=x$. 
Using Eq.~\eqref{eq:g1_from_g0}, similar closed forms can be found for $G_1$ and $p_1$. 

It is enlightening to extract the singular behavior of $G_0$ and $G_1$ at small $z$ directly from asymptotic analysis of the governing equations Eq.~\eqref{eq:g0_cubic_poly} and Eq.~\eqref{eq:g1_from_g0}.
At leading order in $z$ as  $z \to 0$, Eq.~\eqref{eq:g0_cubic_poly} becomes
\begin{align}
    G_0^3 +\frac{\phi-1}{z \phi^2} G_0 = 0
\end{align}
The correct branch in region (ii) diverges as,
\begin{align}
\label{eq:G0_mp_asymp}
    G_0 \sim -i\frac{\sqrt{\phi-1}}{z^{1/2} \phi}
\end{align}
Using  Eq.~\eqref{eq:g1_from_g0}, we find
\begin{align}
\label{eq:G1_mp_asymp}
G_1 &\sim -i \frac{1}{z^{1/2} \sqrt{\phi-1}}
\end{align}
This produces the MP singularity in the DOS.

\paragraph{Multicritical point $\phi=1$---}
For $\phi = 1$, $0 = z_- < z_+$ and most of the exact analysis of $G_0$ from the MP phase carries over.
The support of the DOS is $[0, z_+]$ and $p_0(x)$ is explicitly given by Eq.~\eqref{eq:p0_mp_explicit} with similar expressions for $p_1(x)$.
One can expand these forms to find the modified exponents associated with the multicritical point.
However it is more enlightening to derive them from asymptotic analysis of the governing equations.

To leading order in $1/z$, Eq.~\eqref{eq:g0_cubic_poly} becomes
\begin{align}
    G_0^3 - \frac{1}{z\phi^2} = 0
\end{align}
So that, 
\begin{align}
    G_0 \sim \frac{1}{z^{1/3} \phi^{2/3}}
\end{align}
Moreover, Eq.~\eqref{eq:g1_from_g0} reduces to
\begin{align}
    G_1 = G_0(1-G_0) \sim \frac{-1}{z^{2/3} \phi^{4/3}}
\end{align}

We note that the dominant divergence in the full DOS, $p(x) \propto \phi p_0(x) + p_1(x) \sim x^{-2/3}$ come from the $l=1$ sector.
This is anticipated by the asymptotic forms of $G_0$ and $G_1$ in the MP phase, Eqs.~\eqref{eq:G0_mp_asymp} and \eqref{eq:G1_mp_asymp}, which vanish and diverge respectively as $\phi \to 1^+$.

\paragraph{Gapped phase $\phi < 1$---}
For $\phi < 1$, $0 < z_- < z_+$ and $\Delta < 0$ for (i) $z < 0$ and (ii) $z_- < z < z_+$. 

As before, in region (i), $G_0$ follows the real branch of the 3 roots and there is no contribution to the DOS.

Also as before, the expression Eq.~\eqref{eq:p0_mp_explicit} gives an explicit form for continuous part of the spectrum on the support $[z_-,z_+]$ which turns on with a simple square root singularity at either end. 

Unlike for $\phi \ge 1$, the correct branch of $G_0$ remains real and finite for $z \to 0$,
\begin{align}
    G_0(z) = \frac{1}{\phi - 1} - \frac{z}{ (\phi-1)^4} + O(z^2)
\end{align}
This leads to the pole in $G_1$,
\begin{align}
    {G_1} =\left(\frac{1}{G_0} + \frac{1}{1-\phi G_0}\right)^{-1}
    = \frac{1-\phi}{z} + O(z^{0})
\end{align}
which leads to the delta function in $p(x)$.

\section{Infinite temperature on blockaded chain}

The infinite temperature state on the full chain of length $2L$ has the usual totally mixed density matrix
\begin{align}
    \hat{\rho}^{T=\infty}_{\textrm{TOT}} = \frac{\mathbb{I}}{D} = \frac{1}{D} \sum_{l,r} C_{lr} \delta^l \delta^r
\end{align}
where $D$ is the total Hilbert space dimension.
Unlike in tensor product spaces, the reduced density matrix on the half chain is not simply the identity. Rather,
\begin{align}
    \hat{\rho}^{T=\infty} = \Tr_R \hat{\rho}^{T=\infty}_{\textrm{TOT}} = \frac{1}{D} \sum_{l,r} \delta^l C_{lr} d'_r
\end{align}

Specializing to the blockaded chain, we find,
\begin{align}
	\hat{\rho}^{T=\infty}_{ij} 
	&= \frac{1}{N(\phi^2+2\phi)} \left[ (\phi+1) \delta^0_{ij} + \phi \delta^1_{ij} \right]
\end{align}
The $l=0$ sector has $(\phi+1)/\phi$ times more weight on each state than the $l=1$ sector, due to the additional compatible states in the traced out right half. 

The Renyi and von Neumann entropies follow immediately,
\begin{align}
\label{eq:inft_sn}
	S^{T=\infty}_n &= \frac{-1}{n-1} \ln \Tr (\hat{\rho}^{T=\infty})^n \nonumber\\
	&= \ln N - \frac{1}{n-1}\left[ \ln \left( \frac{\phi(\phi+1)^n + \phi^n}{(\phi^2 + 2 \phi)^n} \right)\right]
\end{align}
and
\begin{align}
\label{eq:inft_s1}
	S^{T=\infty}_1 &= -\Tr \hat{\rho}^{T=\infty} \ln \hat{\rho}^{T=\infty} \nonumber\\
	&= \ln N + \ln(\phi+2) - \frac{\phi+1}{\phi+2}\ln\frac{\phi+1}{\phi}
\end{align}

\end{document}